\documentclass[final]{svjour2}
\usepackage{graphicx}
\usepackage{rotating}
\usepackage{amssymb}
\usepackage{mathptmx}
\usepackage[numbers]{natbib}
\usepackage{sidecap}
\makeatletter
\journalname{Journal of Low Temperature Physics}

\bibpunct{}{}{,}{s}{}{,}

\begin{document}

\newcommand{\hdblarrow}{H\makebox[0.9ex][l]{$\downdownarrows$}-}
\title{High-resolution kaonic-atom x-ray spectroscopy with
transition-edge-sensor microcalorimeters}

\author{S.~Okada \and  D.A.~Bennett \and  W.B.~Doriese \and  J.W.~Fowler
\and K.D.~Irwin \and S.~Ishimoto \and M.~Sato \and  D.R.~Schmidt \and
D.S.~Swetz \and H.~Tatsuno \and  J.N.~Ullom \and S.~Yamada}

\institute{S.~Okada \and M.~Sato \and S.~Yamada\\
RIKEN Nishina Center, RIKEN, Wako, 351-0198, Japan\\
\email{sokada@riken.jp}\\
\\
D.A.~Bennett \and  W.B.~Doriese \and  J.W.~Fowler \and
K.D.~Irwin \and  D.R.~Schmidt \and D.S.~Swetz \and  J.N.~Ullom \\
National Institute of Standards and Technology, Boulder, CO 80303, USA\\
\\
S.~Ishimoto \\
High Energy Accelerator Research Organization, Ibaraki 305-0801, Japan\\
\\
H.~Tatsuno \\
Laboratori Nazionali di Frascati, INFN, I-00044 Frascati, Italy}

\date{26.07.2013}

\maketitle

\begin{abstract}

 We are preparing for an ultra-high resolution x-ray spectroscopy
 of kaonic atoms using an x-ray spectrometer
 based on an array of superconducting transition-edge-sensor
 microcalorimeters developed by NIST.
 The instrument has excellent energy resolutions
 of 2 -- 3 eV (FWHM) at 6 keV and a large collecting area of about 20 mm$^2$.
 This will open new door to investigate kaon-nucleus strong interaction
 and provide new accurate charged-kaon mass value.


\end{abstract}

\section{Introduction}

 Any negatively charged leptons or hadrons
 (e.g., $\mu^-$, $\pi^-$, $K^-$, $\overline{p}$, $\Sigma^-$)
 apart from the conventional electrons
 can be bound by the Coulomb field of an atomic nucleus.
 The Coulomb-bound system, so-called exotic atom,
 is essentially a hydrogen-like atom except for its short lifetime and
 the large reduced mass resulting in unusually small orbital radii.
 In the case of a hadron feeling strong interaction with the atomic
 nucleus, so-called hadronic atom, because of being short-range force,
 the effects appear in the most tightly bound energy level
 being the most perturbed by the strong force
 as a shift from that given only by the electromagnetic interaction,
 and a broadening
 due to absorption of the hadron by the nucleus.
 The shift and width can be experimentally extracted
 from characteristic x-ray-emission spectroscopy of the hadronic-atom
 x-rays feeding the low-lying state.
 This offers a unique opportunity
 to investigate the strong interaction between hadron
 and nucleus (or nucleon) at the low energy limit
 which is unobtainable in typical scattering experiments.
  
 Anti-kaon $\overline{K}$, namely $K^-$ or $\overline{K}^0$,
 is the lightest hadron containing a strange valence quark.
 The low-energy $\overline{K}N$ system has attracted attention as a
 testing ground for chiral SU(3) dynamics in low-energy QCD and
 the role of the interplay between spontaneous and explicit chiral
 symmetry breaking due to the relatively large mass of the strange quark.

 In the simplest kaonic atom, i.e., kaonic hydrogen ($K^-p$),
 the $1s$-atomic-state shift and width
 deduced from the spectroscopy
 of $2p$--$1s$ transition of the $K^-p$ atom x-rays at $\sim$6.5 keV
 are directly related to the real and imaginary parts of
 the complex $K^-p$ $S$-wave scattering length.
 Therefore the data have been intensively collected
 so far \cite{KpX,DEAR,SidKp}.
 The results indicate that
 $\overline{K}N$ interaction in low energy is strongly attractive,
 which leads to particular interests
 in the depth of the real $K^-$-nucleus potential at zero energy
 in connection with possible kaon condensation in astrophysical
 scenarios (e.g., neutron star) and the possible existence of
 ``deeply-bound nuclear $K^-$ states''.

 Many experiments have collected data on a variety of targets
 other than kaonic hydrogen atom so far\cite{Bat97}.
 Despite of the fact
 most of theories reproduce the data \cite{Bat97,Hir00},
 there has been a conflict in the potential depth
 between phenomenological potentials
 and potentials constructed from more fundamental approaches.
 The former are typically 180 MeV deep,
 whereas the latter are less than 50 MeV deep.
 Apart from kaonic-atom experiments,
 a lot of experimental search for the deeply-bound nuclear $K^-$ states
 have been performed in a past decade;
 however, only a small amount of information is available
 \cite{FINUDA,DISTO},
 which is not sufficient to discriminate between a variety of
 conflicting interpretations.

\begin{figure}
\begin{center}
\includegraphics[%
  width=0.84\linewidth,
  keepaspectratio]{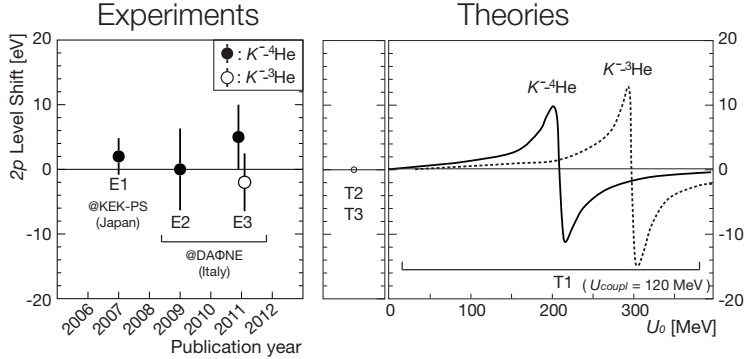}
\end{center}
\vspace*{-3mm} 
\caption{The $2p$-level shift of
 $K^-$-$^3$He and $K^-$-$^4$He atoms
 obtained from the recent experiments (left) at KEK-PS (E1\cite{E570})
 and DA$\rm{\Phi}$NE (E2\cite{SidHe4}, E3\cite{SidHe3})
 compared with the theoretical calculations (right).
 A recent calculation by the use of the coupled-channel model
 as a function of the real part ($U_0$) of the potential depth
 at a coupled potential depth of $U_{coupl}$=120 MeV (T1\cite{Aka05}),
 compared with other calculations showing
 shifts being very close to zero (T2\cite{Bat97},T3\cite{Hir00}).
 Experimental results so far are not accurate enough to distinguish
 between different models.}
\label{Fig:KHeComp}
\end{figure}

 Recently,
 the kaonic-helium atom ($K^-$-He) has attracted interest
 in connection with the theoretical predictions\cite{Bat90,Aka05}
 that a large shift and width of the $2p$ level may appear
 near the resonance between atomic and nuclear poles,
 which is closely related
 to the existence of deeply bound nuclear $K^-$ states.
 A recent calculation\cite{Aka05} shows the shift and width
 as a function of the real part ($U_0$) of the potential depth at
 a certain coupled potential depth ($U_{coupl}$=120 MeV)
 with the different values
 for $K^-$-$^3$He and $K^-$-$^4$He atoms as shown in Fig.\ref{Fig:KHeComp}.
 The calculation was based on the coupled-channel
 approach between the $\overline{K}N$ channel
 and the $\Sigma \pi$ decay channel.
 A large shift of $|\epsilon_{2p}| \sim$ 10 eV and
 width of $|\Gamma_{2p}| \sim$ 20 eV are predicted for $K^-$-$^4$He
 when the potential depth is at around $\sim$ 200 MeV,
 whereas most of theoretical calculations predict
 smaller values, e.g., $\epsilon_{2p} \sim$ -0.2 eV
 and $\Gamma_{2p} \sim$ 2 eV \cite{Bat90}.
 Measurements of $3d$--$2p$ transitions
 in $K^-$-$^3$He (6.2 keV) and $K^-$-$^4$He (6.5 keV)
 have been performed so far with the conventional semiconductor
 spectrometers having the FWHM energy resolution of
 typically $\sim$ 200 eV at 6 keV\cite{E570,SidHe4,SidHe3},
 which is insufficient to see such a small spectral effects due to the
 strong interaction.

\section{A proposed experiment}

 Aiming at a breakthrough to the current situation,
 we are planning to
 perform ultra-high-resolution x-ray spectroscopy of kaonic atoms using
 an array of NIST's transition-edge-sensor (TES)
 microcalorimeters having a FWHM energy resolution of 2 -- 3 eV at 6 keV,
 being two order of magnitude improved resolution
 compared with the conventional semiconductor detector.
 The TES is a thermal sensor that measures an energy deposition by the
 increase of resistance of a superconducting film biased within the
 superconducting-to-normal transition.
 The detailed working principles and the recent progress of the NIST's TES
 system are described in Refs. \cite{Ens05,Ben12}.
 A preliminary result of a recent test measurement shows
 an excellent FWHM time resolution of $\sim$ 0.5 $\mu$sec as well.
 Now we are
 preparing another test measurement at a hadron beamline to evaluate an
 in-beam performance of the x-ray spectrometer.


 We plan to use a NIST-designed 160 pixel TES array
 each having 350 $\mu$m square collecting area per pixel,
 being $\sim $20 mm$^2$ in total,
 together with a time-division SQUID multiplexer readout system.
 For the cryostat,
 we utilize a pulse-tube adiabatic demagnetization refrigerator (ADR)
 developed by NIST and HPD, 102 DENALI\cite{HPD},
 whose base temperature is 50 mK.
 The size of this system is 33 cm $\times$ 22 cm $\times$ 66 cm tall
 which is relatively portable and compact.
 The portability is essential for our experiment constrained by very
 limited and busy beam time.

 The experiment will be performed at a kaon beamline
 at J-PARC hadron hall, e.g., K1.8BR beamline \cite{K1.8BR}.
 Kaons are produced by bombarding the production target (Ni, Ag or Pt)
 with the primary proton beam
 from J-PARC 50 GeV proton synchrotron,
 and extracted through the kaon beamline.
 The kaonic-helium atoms are created by stopping kaons
 in a helium target.

\begin{figure}
\begin{center}
\includegraphics[%
  width=0.90\linewidth,
  keepaspectratio]{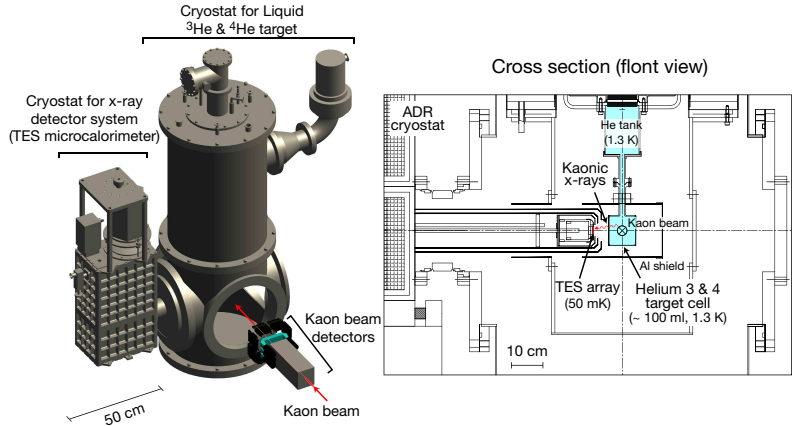}
\end{center}
\vspace*{-3mm} 
\caption{(Color online) A schematic birds-eye view (left) and
 a cross section front view (right)
 of a possible experimental setup around the experimental target.
 A set of beam detectors,
 a cryogenic liquid-$^3$He and $^4$He target system,
 and a NIST's TES x-ray spectrometer system are shown.
 }
\label{Fig:setup}
\end{figure}

 Fig.\ref{Fig:setup} shows a possible 
 a possible experimental setup.
 Incident kaons extracted with a momentum of $\sim$ 900 MeV/$c$ are
 degraded in carbon degraders, 
 counted with beamline counters
 together with Lucite $\check{\mbox{C}}$herenkov counters
 for reducing the pion contamination,
 tracked by a high-rate beamline drift chamber
 and stopped inside the liquid-helium target.
 To ensure that kaons are stopped in the target,
 an anti-coincidence counter
 is installed just downstream side of the target.
 X-rays emitted from the $K^-$-He atoms are detected
 by a TES spectrometer viewing the target from a side.
 
 The main trigger will be defined by 
 the beamline counters to identify incoming kaons
 and the anti-coincidence counter to ensure stopped kaons
 in the target.
 X-ray signals of TES are coincident with the trigger
 to reduce accidental backgrounds.

\begin{SCfigure}
  \centering
 \includegraphics[%
 width=0.60\linewidth,
 keepaspectratio]{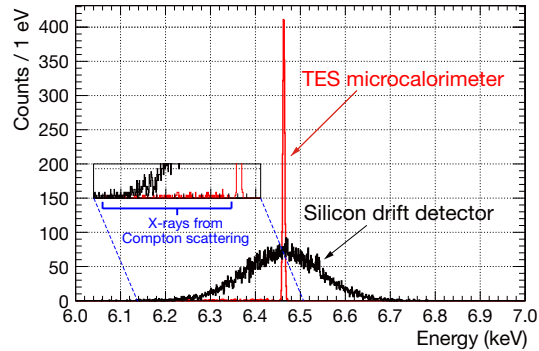}
 \caption{(Color online) Simulated spectra of
 of the $3d$--$2p$ transitions of kaonic-$^4$He atoms
 detected by different x-ray detectors:
 a conventional silicon drift detector
 with a FWHM energy resolution of 190 eV at the energy,
 and a TES microcalorimeter with a FWHM energy resolution of 5 eV.
 }
 \label{Fig:spectrum}
\end{SCfigure}

 
 Fig.\ref{Fig:spectrum} is a result of a GEANT4 simulation
 where $K^-$-$^4$He $3d$--$2p$ x-rays
 are measured by different two x-ray detectors.
 One is a typical conventional silicon drift detector
 used in the previous experiments\cite{E570,SidHe4,SidHe3}
 having a 400-$\mu$m-thick active layer
 with 100 mm$^2$ collecting area
 and 190 eV (FWHM) energy resolution.
 The other is a TES spectrometer
 having 5-$\mu$m-thick Bi absorbers
 with 20 mm$^2$ collecting area
 and 5 eV (FWHM) energy resolution.

 In the previous experiment\cite{E570},
 there has been a non-negligible effect of spectral distortion
 due to Compton scattering in the liquid $^4$He target.
 As shown in Fig.\ref{Fig:spectrum},
 the effect is completely separated from the distinct x-ray peak
 measured by TES.
 Although the other experiments\cite{SidHe4,SidHe3} utilised a gaseous helium
 target so as to reduce the effect, we can now use a liquid helium
 target to efficiently stop kaons without suffering the Compton
 scattering problem.

 A lot of fluorescence x-rays
 (e.g., Fe K$_{\alpha1}$ 6.40 keV, Fe K${_{\alpha2}}$ 6.39 keV
 and Mn K$_{\beta}$ 6.49 keV) and kaonic x-rays other than $K^-$-He x-rays
 were possible serious background sources in the past experiments.
 In reversal, these are rather useful as good calibration lines
 in the proposed measurement with a TES spectrometer.
 The precise energy calibration is essential in such a high resolution
 measurement; and thus we will carefully select our calibration sources
 of the fluorescence x-rays whose energies, width and satellite peaks
 are well known.
 Moreover a simultaneous measurement of $3d$--$2p$ x-rays
 in $K^-$-$^4$He (6.5 keV) and $K^-$-$^3$He (6.2 keV) atoms
 with a mixed liquid $^3$He and $^4$He target is possible,
 which will drastically reduces the systematic errors
 of relative x-ray energies and widths for both atoms.

 Referring the yield of $K^-$-$^4$He $3d$--$2p$ x-rays
 obtained in the previous experiment\cite{E570},
 the yield was roughly estimated to be about 30 events per day
 assuming that the J-PARC K1.8BR beamline is employed
 with an operational proton beam power of 60 kW.
 Four-days data acquisition will give about 120 events
 resulting in a statistical error of 0.1 eV
 in x-ray energy,
 assuming 2 -- 3 eV (FWHM) energy resolution without background.
 Thus, we could obtain more than one order of magnitude improved
 statistical errors of the x-ray energy and the width values
 compared to the previous experiments
 with a reasonable data acquisition time.

 As well as $K^-$-He atoms,
 there is another theoretical suggestion
 of a precise measurement of $3d$--$2p$ x-rays
 in $K^-$-$^6$Li and $K^-$-$^7$Li  ($\sim$ 15 keV),
 whose result has a possibility to discriminate between deep
 (phenomenological) or shallow (chiral) potentials\cite{HirPri}.
 Moreover there are other remarkable studies using
 ultra-high-resolution kaonic x-ray spectroscopies
 with TES spectrometers we intend to perform as follows:
 
\begin{enumerate}
 \item Charged kaon mass measurement:
       Kaonic-atom x-ray spectroscopy has been also
       utilized as a tool for
       measuring the charged kaon mass.
       The latest value
       is determined by the average of the six measurements
       to be 493.677 $\pm$ 0.013 MeV (S=2.4)\cite{PDG}.
       This relatively large error mainly comes from
       two data whose central values differ vastly about 60 keV;
       thus a new accurate measurement has been eagerly awaited.
       We aim to improve the precision of this mass measurement
       with a high-precision measurement
       of $K^-$-$^{12}$C $5$--$4$ x-ray ($\sim$ 10 keV).
       The precision is expected to be $\sim$ 2.5 keV corresponding to the
       accuracy of $\sim$ 0.05 eV for the x-ray energy.

\vspace{1.5mm}
       
 \item Study of $K^-$ multi-nucleon absorption process:
       Recent theoretical studies
       of strong interaction effects in kaonic atoms
       suggest that analysing
       so-called 'lower' and 'upper' levels in the same atom
       could separate one-nucleon absorption from
       multinucleon processes\cite{Fri13}.
       For this study,
       direct measurements of upper level widths in
       addition to lower level widths are essential
       for medium-weight and heavy kaonic atoms
       (100 -- 450 keV x-rays)
       whose candidates were examined recently\cite{FriOka}.
       The measurements will be performed in future
       with a gamma-ray TES spectrometer
       similar to what NIST has already achieved,
       having a 256 pixel array resulting in 5 cm$^2$ collecting area
       and an averaged FWHM energy resolution of 53 eV
       at 100 keV\cite{Ben12}.
\end{enumerate}

\section{Conclusions}

 We are preparing a next-generation kaonic-atom x-ray spectroscopy
 using an x-ray spectrometer
 based on an array of superconducting
 TES microcalorimeters developed by NIST.
 Precise measurements of $3d$--$2p$ x-rays
 in kaonic-helium atoms ($\sim$ 6 keV)
 and kaonic-lithium atoms ($\sim$ 15 keV)
 with two order of magnitude improved energy resolution
 compared with the conventional silicon detectors
 will enable us to see small spectral effects due to the
 strong interaction.
 This will add a fresh dimension to
 the study of the depth of the real $K^-$-nucleus potential at zero energy
 in connection with the existence of deeply-bound nuclear $K^-$ states.
 Additionally, 
 we intend to improve the precision of the charged kaon mass measurement
 with high-precision measurement
 of $K^-$-$^{12}$C $5$--$4$ x-ray ($\sim$ 10 keV),
 and study the $K^-$ multi-nucleon absorption process in future
 with direct measurements of upper level widths
 in medium-weight and heavy kaonic atoms
 using a multi-pixel TES spectrometer 
 with a useable dynamic range above 400 keV.

\begin{acknowledgements}
 This work was supported by JSPS KAKENHI Grant Number 25105514
 and by Incentive Research Grant from RIKEN.
\end{acknowledgements}


\end{document}